\documentclass[namedreferences]{SolarPhysics}
\usepackage[optionalrh]{spr-sola-addons} % For Solar Physics 
\usepackage[font=small]{caption}
\usepackage{graphicx}                    % For eps figures, newer & more powerfull
\usepackage{booktabs}
\usepackage{lscape}
\usepackage{threeparttable}
\usepackage{color}                       % For color text: \color command
\usepackage{url}                         % For breaking URLs easily trough lines
\newcommand{\apj}{    {\it Astrophys. J.}}

\begin{document}

\begin{article}

\begin{opening}

\title{No Evidence Supporting Flare Driven High-Frequency Global Oscillations}

%%%%%%%%%%%%%%%%%%%%%%%%%%%%%%%%%%%%%%%%%%%%%%%%%%%
%% Authors Names
%
\author{M.~\surname{Richardson}$^{1}$\sep
        F.~\surname{Hill}$^{2}$\sep
        K. G. ~\surname{Stassun}$^{1}$      
       }

%%%%%%%%%%%%%%%%%%%%%%%%%%%%%%%%%%%%%%%%%%%%%%%%%%%
%% Runningheads
%
%\runningauthor{}
%\runningtitle{}

%Affilations 
  \institute{$^{1}$ Vanderbilt University\\ %email: \url{matthew.d.richardson.1@vanderbilt.edu} email:\url{keivan.stassun@vanderbilt.edu}\\ 
             $^{2}$ National Solar Observatory %email: \url{hill@noao.edu} \\
             }

%Abstract 
\begin{abstract}
The underlying physics that generates the excitations in the global low-frequency, $<$ 5.3 mHz, solar acoustic power spectrum is a well known process that is attributed to solar convection; However, a definitive explanation as to what causes excitations in the high-frequency regime, $>$ 5.3 mHz, has yet to be found. \citeauthor{Karoff08} (\apj{} \textbf{678}, 73-76, \citeyear{Karoff08}) concluded that there is a correlation between solar flares and the global high-frequency solar acoustic waves. We have used the Global Oscillations Network Group (GONG) helioseismic data in an attempt to verify \citeauthor{Karoff08} (\citeyear{Karoff08}) results as well as compare the post-flare acoustic power spectrum to the pre-flare acoustic power spectrum for 31 solar flares. Among the 31 flares analyzed, we observe that a decrease in acoustic power after the solar flare is just as likely as an increase. Furthermore, while we do observe variations in acoustic power that are most likely associated with the usual $p$-modes associated with solar convection, these variations do not show any significant temporal association with flares. We find no evidence that consistently supports flare driven high-frequency waves. 
\end{abstract}

%Keywords
\end{opening}
%-------------------------------------------------

% Sections
\section{Introduction}\label{Intro} 
Before the advent of helioseismology in the 1970s, the study of the internal structure and dynamics of the Sun was only possible with one-dimensional theoretical models. While the subsequent study of  solar oscillations has helped answer many questions pertaining to the solar interior, it has also presented new questions. One of these questions is whether the oscillations can be influenced by flares. \citeauthor{Wolff72} (\citeyear{Wolff72}) was probably the first to suggest that large solar flares could excite free modes of oscillation of the entire Sun. Since then, much research has been conducted to determine if solar flares can affect the global oscillations, and in particular, the high-frequency waves.

One of the first reports of the effects of solar flares on solar oscillations was published by \citeauthor{Haber88} (\citeyear{Haber88}). Dopplergrams of intermediate-degree $p$-modes were used to observe the change in acoustic power before and after a major solar flare event. They concluded that the flare caused an average increase in power the day after the flare. Shortly after, \citeauthor{Braun90} (\citeyear{Braun90}) analyzed  the interaction between the $p$-mode oscillations and an X-class flare in NOAA (National Oceanic and Atmospheric Administration) active region 5359 using radial velocity observations obtained from the Kitt Peak vacuum telescope. Their results indicated no evidence of an increase in the oscillation amplitude correlated with the flare event. 
\citeauthor{Foglizzo98} (\citeyear{Foglizzo98}) studied the correlations between low-degree modes to determine if they could be excited by mechanisms other than granulation. It was concluded that X-class solar flares most likely do not excite low-degree oscillations due to inconsistencies in energy conservation. 
\citeauthor{Ambastha06} (\citeyear{Ambastha06}) search for a correlation between the power of low-degree modes observed in velocity and flare energy release was generally inconclusive. \citeauthor{Gavryusev99} (\citeyear{Gavryusev99}) analysis found a high correlation between temporal changes in low-degree spectral power and coronal mass ejections, which suggests that there exists a relationship between p-mode power and solar activity. Although, it is unclear whether thier finding can be extended to solar flares.
\citeauthor{Donea05} (\citeyear{Donea05}) found that a helioseismic holography analysis of the seismic emission from two X-class solar flares indicates that the hard X-ray emission produced during solar flares are associated with acoustic signatures.  
\citeauthor{Kumar06} (\citeyear{Kumar06}) reported an enhancement in power maps at frequencies from 5 mHz to 6.5 mHz of a solar active region in which a large X-class flare was produced. Finally, Kosovichev has observationally shown that flares can excite local waves at the location of the flare (\citeauthor{Kosovichev98} \citeyear{Kosovichev98}; \citeauthor{Kosovichev06} \citeyear{Kosovichev06}, \citeauthor{Kosovichev11} \citeyear{Kosovichev11}).

Thus, while it is established that large flares can generate local acoustic waves in the solar atmosphere, there is still conflicting evidence for the role of flares in affecting global oscillations. In particular, \citeauthor{Karoff08} (\citeyear{Karoff08}), hereafter KK, used solar disk-integrated intensity data from the Variability of Solar IRradiance and Gravity (VIRGO) instrument, on-board the Solar and Heliospheric Observatory (SOHO) satellite, and solar X-ray flux data from the Geostationary Operational Environmental Satellites (GOES) to study the effects of flares on the global oscillations.  They found a high correlation between the power in the high-frequency acoustic power spectrum and variations in the solar X-ray flux. 
Kumar and collaborators have found evidence supporting the theory of flare driven high-frequency acoustic oscillations (\citeauthor{Kumar09} \citeyear{Kumar09}; \citeauthor{Kumar10} \citeyear{Kumar10}). Using a wavelet analysis of Michelson Doppler Imager (MDI) and Global Oscillation at Low Frequency (GOLF) data, they found enhancements of high-frequency low-degree power associated with three powerful flares in the MDI data, but only marginally in the GOLF data. Both of these instruments are also on SOHO. In addition, \citeauthor{Kumar11} (\citeyear{Kumar11}) reported a significant post-flare enhancement in the high-frequency global velocity oscillations for both GOLF and the Global Oscillations Network Group (GONG) velocity data of a X-class solar flare in NOAA active region 10930.
Most recently, \citeauthor{Chakraborty11} (\citeyear{Chakraborty11}) were able to reproduce the KK results using acoustic data obtained by the Sun PhotoMeter (SPM) instrument on SOHO and GOES X-ray data obtained by the Solar X-ray Imager. However, the results they obtained from correlating the GOES X-ray data with the acoustic data produced by GONG instruments showed a weaker correlation between solar flares and acoustic power at frequencies higher than the solar acoustic cut-off frequency.

In this paper we present an analysis using GONG and GOES data to determine if global helioseismic signatures are affected by major solar flare activity. The analysis was conducted using two independent methods. The first method was a comparative analysis between pre-flare and post-flare acoustic power spectra for a few selected major solar flares. The objective of this portion of the analysis was to see if there were increases in the power spectra directly after the flare event as Wolff's (\citeyear{Wolff72}) hypothesis suggested. The second method was an attempt to reproduce the KK results by constructing frequency vs time plots for four different ranges of spherical harmonic degree and comparing those results to temporal variations in the solar X-ray flux. Given that space-based detectors are constantly being bombarded by energetic solar particles from events like solar flares and coronal mass ejections, it is possible that the KK data set may have been affected by such particle events, which could influence the data.  As GONG data are ground-based solar observations and can not be affected by such interference, the GONG observations are an ideal data set to use to test the results obtained by \citeauthor{Karoff08} (\citeyear{Karoff08}). 

\section{Data and Analysis Methods}\label{Data/Analysis Methods} 
\noindent In this study we examine approximately 14 years, from Dec. 9, 1995 to April 29, 2010, of GONG global helioseismic observations. The GONG instruments generate polarized disk intensity images of the Sun, taken every 60 seconds, as their primary data product. After reductions and calibrations, the intensity images are converted to Doppler images. Each Doppler image from each of the six GONG sites is decomposed into a set of spherical harmonic amplitudes or coefficients ranging in degree $\ell$ from 0 to 200 (\citeauthor{Hill96}, \citeyear{Hill96}). The coefficients from simultaneous measurements are merged into a set of coherent time series for each individual value of $\ell$ and $|m|$, where $m$ is the azimuthal degree, with $0 \le |m| \le \ell$. The spherical harmonic time series data show the temporal behavior of global solar $p$-modes. 

The time series are converted into power spectra by  taking the FFT of the corresponding spherical harmonic coefficient time series. This produces spectra as a function of $m$ and $\nu$ for each $\ell$, with $ -\ell \le m \le \ell$. This step also separates the modes in frequency $\nu$ with identical $(\ell,m)$ but different $n$, where $n$ is the radial order of the mode. The azimuthal order $m$ yields information about the phase and spatial pattern of a $p$-mode on the solar surface. However due to solar differential rotation,  multiplet $p$-modes with identical $(n, \ell)$ but different $m$  that would have identical frequencies if the solar surface rotated as a solid body, acquire non-degenerate frequencies with respect to azimuthal order $m$. Note that, since we cannot observe the entire spherical surface of the Sun, a power spectrum computed for a specific $\ell$ also contains power from several adjacent values of $\ell$, producing additional features in the spectrum. These artifacts are known as spatial leaks (\citeauthor{Howe98}, \citeyear{Howe98}).

In order to decrease the noise in the spectra generated by the stochastic excitation of the modes, we next shift the spectrum for each $(\ell,n)$ multiplet by a frequency $\delta \nu$ as a function of $m$  by the amount $\delta \nu = m P_{(\ell,n)}$, where $P_{(\ell,n)}$ is a Legendre polynomial. This removes the frequency shift arising from differential rotation and other effects. Next, the spectrum is averaged over $m$, and then over four ranges of $\ell$, as follows: $\ell_1$: $0 \le \ell \le 50$; $\ell_2$: $51 \le \ell \le 100$;  $\ell_3$: $101 \le \ell \le 150$; and $\ell_4$: $151 \le \ell \le 200$. These $m$- and $\ell$-averaged spectra are computed over a variety of time spans, as described below.

We used the soft channel (1-8 \r{A}) flux measured by the X-ray instruments on the GOES satellites to compare the occurrence and strength of major solar flares to increases in solar acoustic power. This time series is averaged over the same time intervals as the oscillation data for comparisons. Rapid changes in the X-ray flux are used to identify the times of flare maximum.

\subsection{Pre-flare/Post-flare Comparison Analysis}\label{Pre-flare/Post-flare Comparison Analysis} 
We compared oscillation power spectra before and after individual large flares to test Wolff's hypothesis (\citeyear{Wolff72}) that flares can stimulate the normal modes of the Sun.
We selected 31 flares that were of M5-class or higher and within $45^\circ$ of the center of the solar disk.  For each flare, we calculated pre-event and post-event average power spectra of one day duration, divided at the time of the maximum intensity of the flare. 
The power ratio of the post-flare to pre-flare power spectra was then calculated as a function of $\nu$ in the four ranges of $\ell$ described above. As a control, the same procedure was performed for points in time when major solar flares, M5-class or higher, were not present. 

\subsection{Production of power as a function of frequency and time}\label{Production of ft plots}
For confirming the KK results, we computed power as a function of frequency and time in a manner similar to the method described in \citeauthor{Karoff08} (\citeyear{Karoff08}), but with a few differences. Consecutive 7.5-day $m$-averaged, $\ell$-averaged FFT power spectra were calculated as explained above. We chose 7.5-day time strings so that the frequency resolution of our analysis was in agreement with the frequency resolution of the KK analysis. Each time string was displaced 18 hours from the start time of the preceding time string and the power spectra were then vertically stacked to produce the plots. KK's analysis employed smoothing of the power spectra. No smoothing was employed in our analysis so as to preserve the full rich information content of the data especially for the high frequencies of interest here. This procedure was performed for the four different ranges of $\ell$. In addition, we further averaged these data over frequency and over time to examine the  overall variations of power in time and frequency, respectively.

For consistency, the X-ray data were processed in the same manner as the power spectra. The X-ray time series was averaged in 7.5 day time strings in which the starting time of each interval was shifted by 18 hours from the starting time of the previous time string. No smoothing was applied to the X-ray data.

\section{Results}

\subsection{Pre-flare/Post-flare Comparison}
\begin{figure}[h!]
\centerline{
\includegraphics[width=5.0 in]{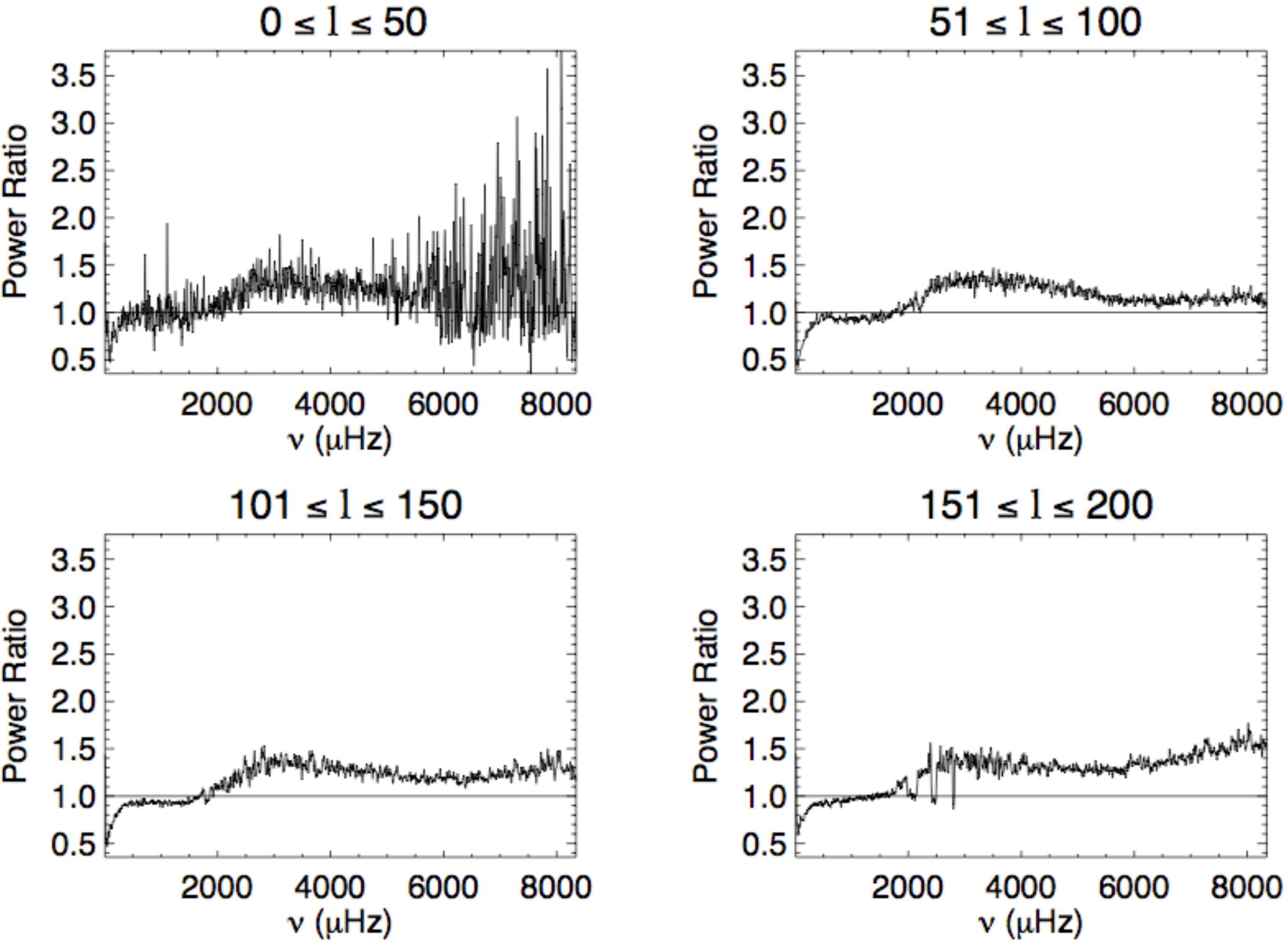}
}
\caption{The power ratio of $m$-averaged,  $\ell$-averaged temporal power spectra before and after the X5.7 flare that occurred on July 14, 2000. The ratios are in the sense of post-flare power divided by pre-flare power, so a ratio greater than 1 indicates a power enhancement.}
\label{Fig: 1}
\end{figure}

Given the number of flares used for this portion of our analysis, we present plots for only three X class solar flares selected for our comparison analysis. Results obtained for the entire set of flares are summarized in Table 1 and graphically displayed in Figures~\ref{Fig: 5} and~\ref{Fig: 6}. In Figure~\ref{Fig: 1}, for the X5.7 flare of July 14, 2000 (the Bastille Day flare), we see a general increase in acoustic power after the flare for all four ranges of $\ell$ used for this analysis, and for frequencies above about 2 mHz. For this case, Wolff's hypothesis could be considered to be supported. 

\begin{figure}[h!]
\centerline{
\includegraphics[width=5.0 in]{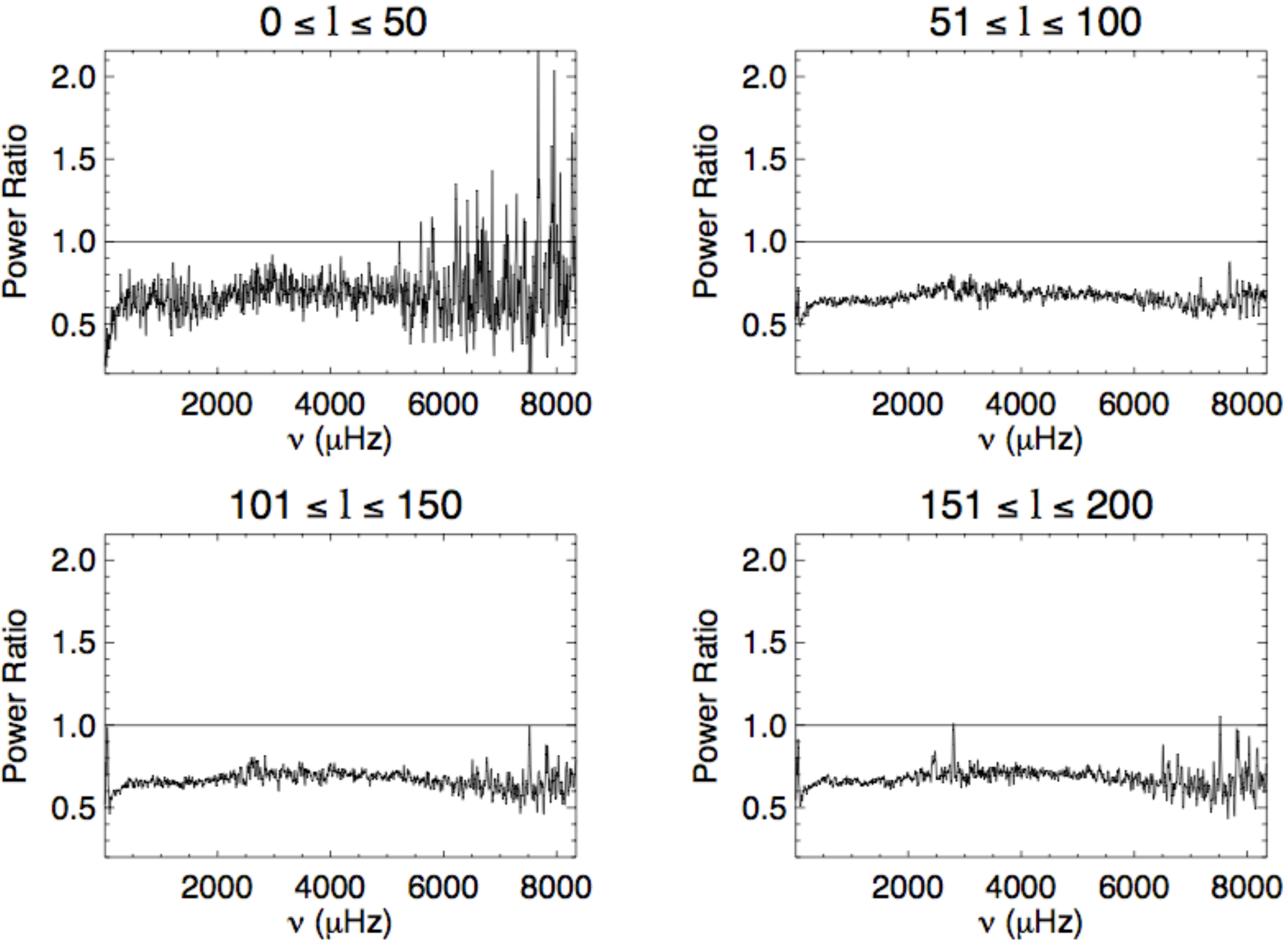}
}
\caption{As for Figure 1, but for the X6.2 flare of Dec. 13, 2001}
\label{Fig: 2}
\end{figure}

\begin{figure}[h!]
\centerline{
\includegraphics[width=5.0 in]{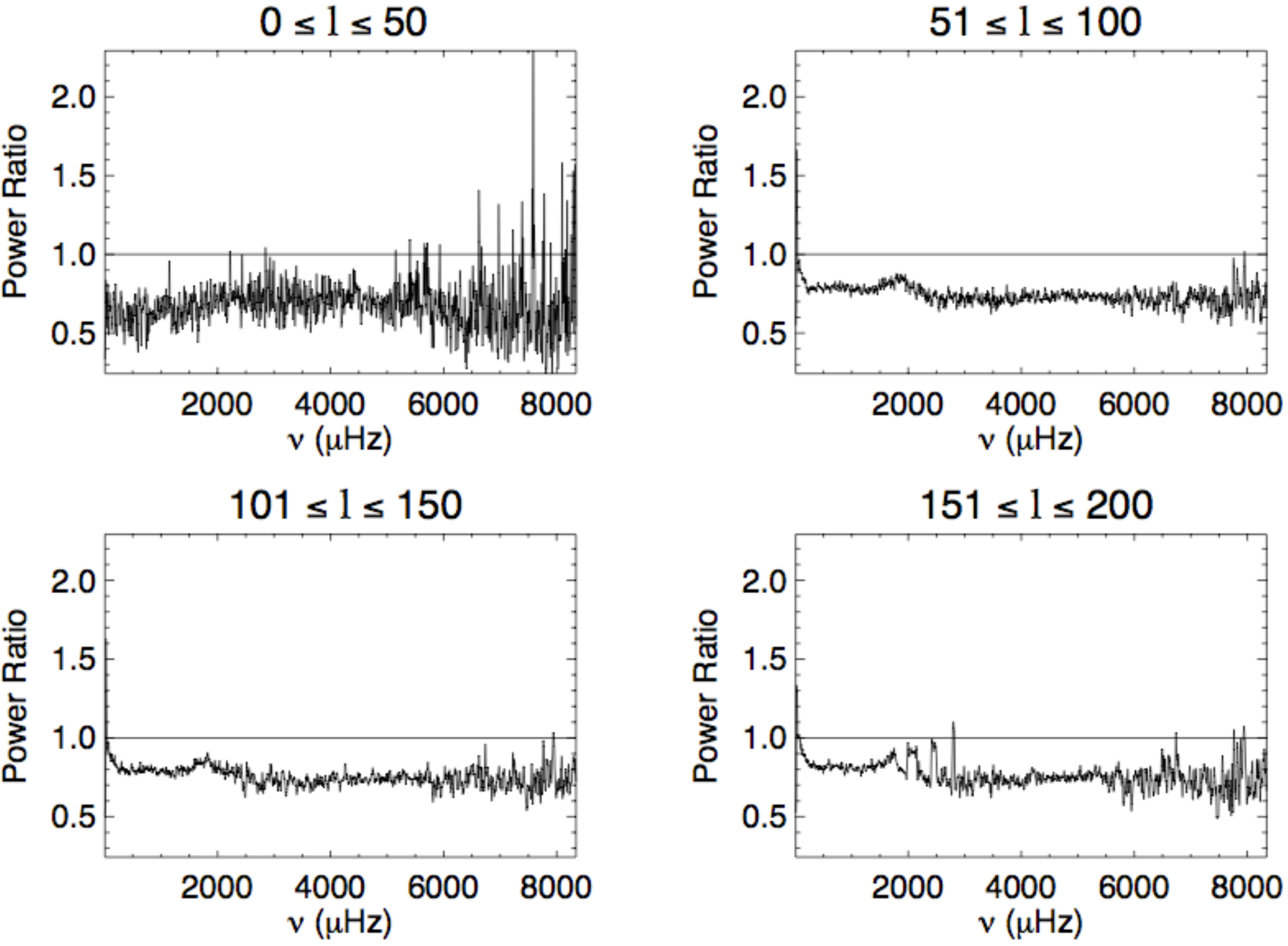}
}
\caption{As for Figure 1, but for the X17.2  flare of Oct. 28, 2003}
\label{Fig: 3}
\end{figure}

\begin{figure}[h!]
\centerline{
\includegraphics[width=5.0 in]{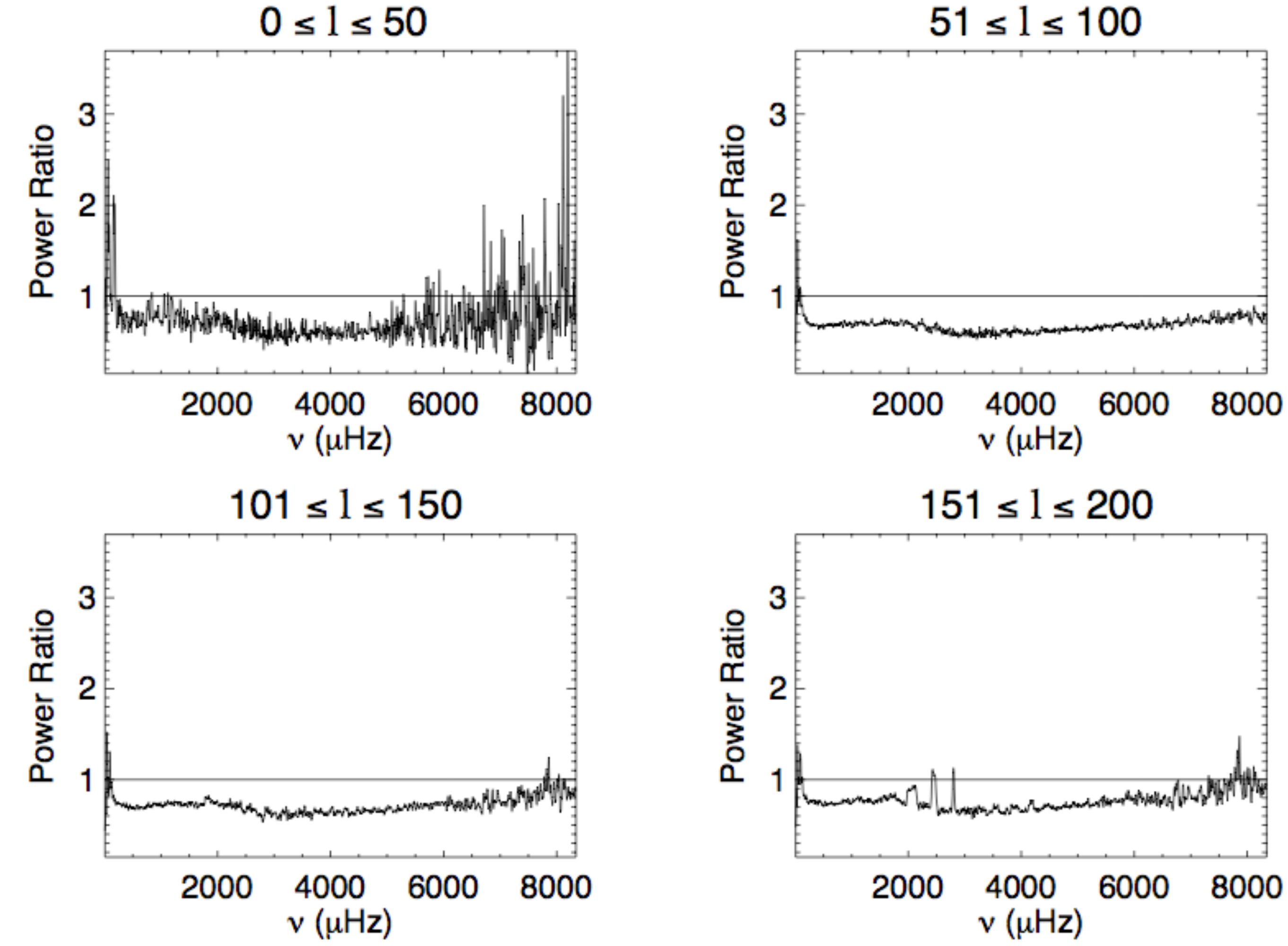}
}
\caption{As for Figure 1, but for an arbitrary quiet period on April 3, 2004.}
\label{Fig: 4}
\end{figure}

\begin{figure}[h!]
\centerline{
\includegraphics[width=5.0 in]{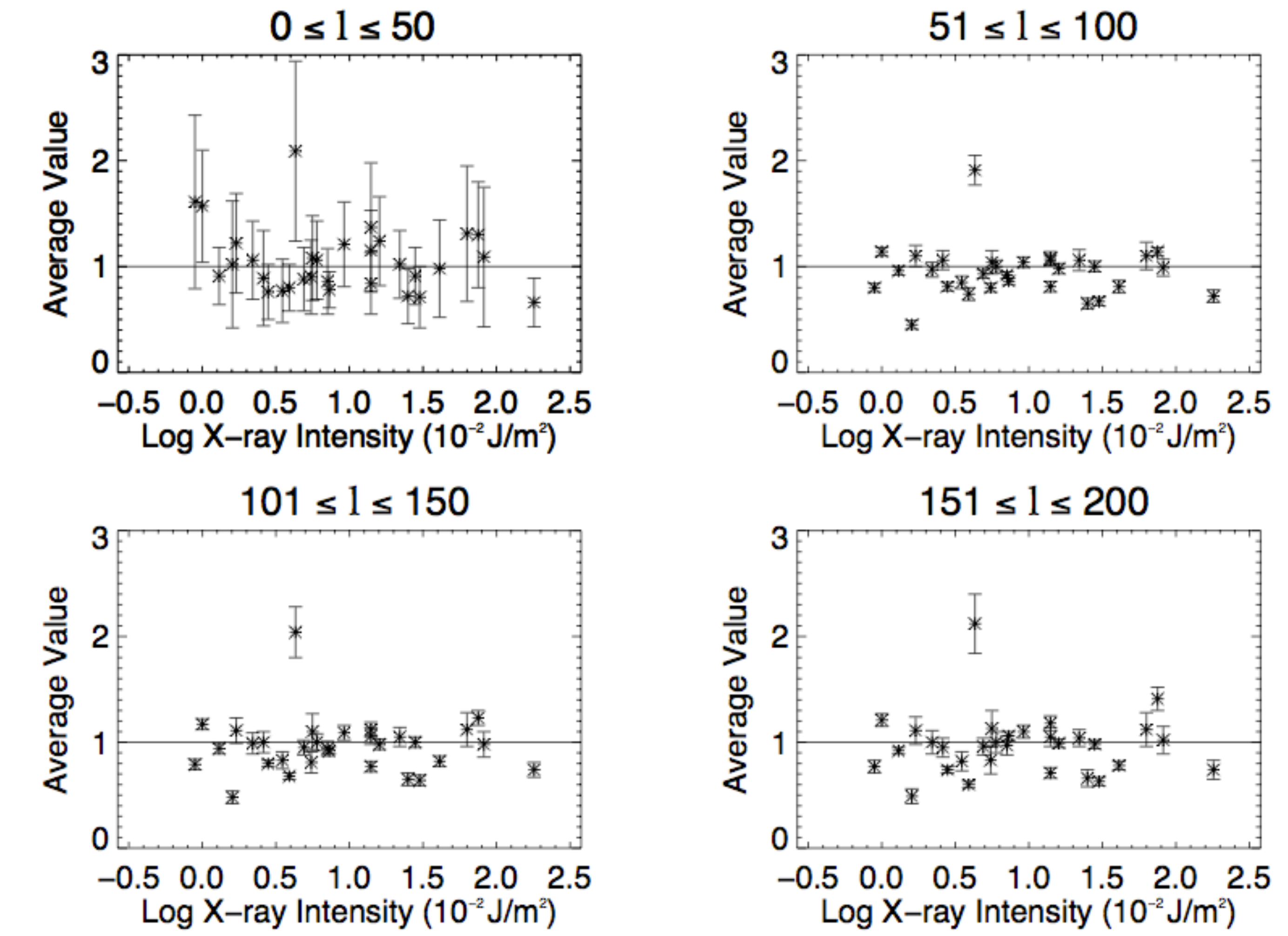}
}
\caption{The average value of the power ratio above the acoustic cutoff frequency ($>$ 5.3 mHz), as seen in columns 4 - 7 in Table 1, plotted as a function of solar X-ray intensity. The error bars are the corresponding standard deviations in columns 8 - 11 in Table 1.}
\label{Fig: 5}
\end{figure}

\begin{figure}[h!]
\centerline{
\includegraphics[width=5.0 in]{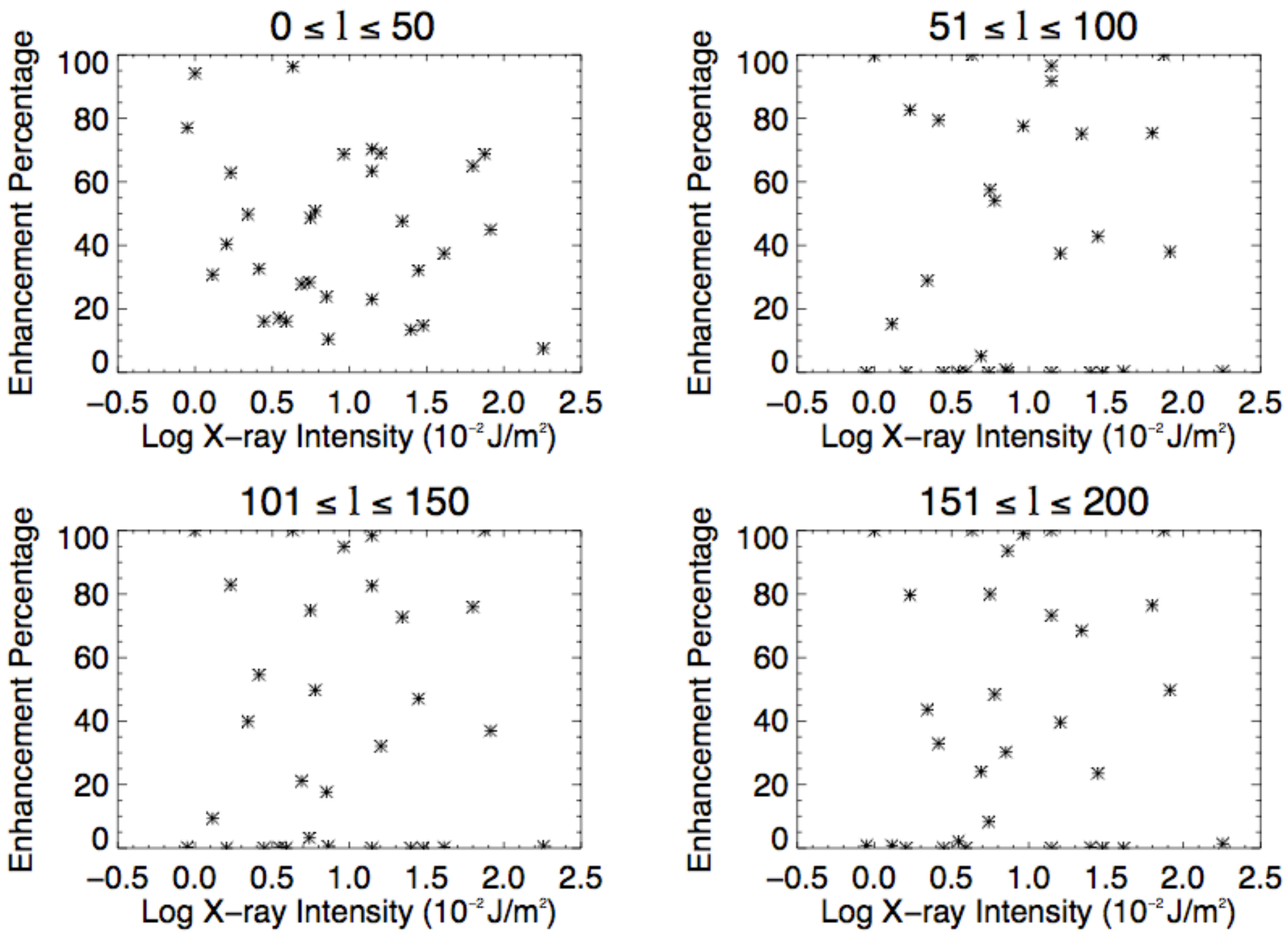}
}
\caption{Percentage of values greater than unity above the acoustic cutoff frequency ($>$ 5.3 mHz), as seen in columns 12 - 15 in Table 1, plotted as a function of solar X-ray intensity.}
\label{Fig: 6}
\end{figure}

However, in Figures~\ref{Fig: 2} and~\ref{Fig: 3} the plots generally show a decrease in power, for all ranges of $\ell$ used, after the occurrence of the flares. Both of these flares, an X6.2 on December 13, 2001, and the Halloween X17.2 event of October 28, 2003, were substantially more powerful than the Bastille Day flare of Figure 1, but show a relative decrease in power rather than an increase. Of the 31 solar flare events selected for our analysis,  as shown in Table 1, 16 (52\%) have an average power ratio above 1 in the $\ell_1$ range; 13 (42\%) have an average power ratio above 1 in the $\ell_2$ and $\ell_3$ ranges, and 14 (45\%) have an average power ratio above 1 in the $\ell_4$ range. The plots in Figure ~\ref{Fig: 4} are for a control quiet time, and show an overall decrease in power similar to that in Figures~\ref{Fig: 2} and~\ref{Fig: 3}. From these results, we can conclude that flare-related variations are probably no different from the variations during quiet times that arise from the stochastic driving of the modes. Furthermore, we calculated the correlation for the data presented in Figures~\ref{Fig: 5} and~\ref{Fig: 6} using both Spearman and Pearson correlation statistics for all four of the $\ell$ ranges plotted. The Spearman rank correlation yielded correlation values less than 60\% confidence and the Pearson correlation values strongly indicate no correlation.

There is a clear increase in the variability of the power ratio above the acoustic cutoff frequency around 5.3 mHz for the lowest range $\ell_1$ $(0 \le \ell \le 50)$. This excess variability is observed for almost all of the 1-day spectral ratios, both for flares and for the quiet Sun. In addition, the high-frequency power ratio in the $\ell_1$ range is always much more variable than at the higher degrees, as shown in Figure~\ref{Fig: 5}. While this suggests that there may be a difference in the excitation and damping of the lowest degree modes, there are also far fewer values of $m$ available for  averaging at the lower degrees. Thus, the higher variation with respect to frequency for the lowest range $\ell_{1}$ may simply be the result of incomplete noise cancellation. However, the variability in the power ratios for all three of the higher ranges of $\ell$ do not show an increasing reduction that would be expected if the effect is mainly due to $m$ averaging.

\begin{landscape}
\begin{table}[p]\tiny
\centering
\begin{threeparttable}[b]
\caption{}
\begin{tabular}{lrrrrrrrrrrrrrrrrrrr}
\toprule
\multicolumn{3}{c}{} & \multicolumn{12}{c}{High-Frequency}\\
\multicolumn{3}{c}{Flare} & \multicolumn{4}{c}{$\mu_{\ell}$\tnote{a}} & \multicolumn{4}{c}{$\sigma_{\ell}$\tnote{b}} & \multicolumn{4}{c}{\scriptsize Enhancement\tnote{c}}\\
\multicolumn{1}{c}{Class} & \multicolumn{1}{c}{Date} & \multicolumn{1}{c}{\scriptsize X-ray Intensity} & \multicolumn{1}{c}{$\ell_1$\tnote{d}} & \multicolumn{1}{c}{$\ell_2$\tnote{d}} & \multicolumn{1}{c}{$\ell_3$\tnote{d}} & \multicolumn{1}{c}{$\ell_4$\tnote{d}} & \multicolumn{1}{c}{$\ell_1$} & \multicolumn{1}{c}{$\ell_2$} & \multicolumn{1}{c}{$\ell_3$} & \multicolumn{1}{c}{$\ell_4$} & \multicolumn{1}{c}{$\ell_1$} & \multicolumn{1}{c}{$\ell_2$} & \multicolumn{1}{c}{$\ell_3$} & \multicolumn{1}{c}{$\ell_4$}\\ 
\multicolumn{2}{c}{} & \multicolumn{1}{c}{(10$^{-2}$ J/m$^{2}$)} & \multicolumn{12}{c}{} \\
\midrule
M 5.0 & July 20, 2000 & 1.00 & 1.57 & 1.14 & 1.17 & 1.21 & 0.53 & 0.04 & 0.05 & 0.06 & 94.12\% & 99.73\% & 100.00\% & 100.00\%\\ 
M 5.1 & April 23, 2003 & 3.50 & 0.77 & 0.85 & 0.83 & 0.82 & 0.30 & 0.06 & 0.08 & 0.09 & 17.11\% & 0.00\% & 0.27\% &  2.14\%\\ 
M 5.2 & June 10, 2000 & 7.30 &  0.78 & 0.86 & 0.92 & 1.06 & 0.17 & 0.03 & 0.03 & 0.04 & 10.43\% & 0.00\% &0.53\% & 93.58\%\\ 
M 5.7 & March 8, 2001 & 1.30 & 0.91 & 0.96 & 0.94 & 0.92 & 0.27 & 0.04 & 0.05 & 0.03 & 30.75\% & 15.24\% & 9.36\% & 0.80\%\\ 
M 5.9 & Oct. 5, 2002 & 4.30 & 2.09 & 1.91 & 2.04 & 2.12 & 0.85 & 0.14 & 0.24 & 0.28 & 96.26\% & 100.00\% & 100.00\% & 100.00\%\\ 
M 6.0 & Sep. 5, 2001 & 1.60 & 1.02 & 0.45 & 0.48 & 0.49 & 0.60 & 0.04 & 0.06 & 0.07 & 40.37\% & 0.00\% & 0.00\% & 0.00\%\\ 
M 6.1 & June 3, 2000 & 3.90 & 0.80 & 0.74 & 0.68 & 0.60 & 0.22 & 0.06 & 0.03 & 0.03 & 16.04\% & 0.27\% & 0.00\% & 0.00\%\\ 
M 6.5 & Oct. 23, 2001 & 5.50 & 0.90 & 0.80 & 0.81 & 0.83 & 0.35 & 0.04 & 0.10 & 0.13 & 28.34\% & 0.00\% & 3.21\% & 8.29\%\\ 
M 6.8 & Dec. 20, 2002 & 0.89 & 1.61 & 0.80 & 0.79 & 0.77 & 0.82 & 0.04 & 0.05 & 0.06 & 77.01\% & 0.00\% & 0.27\% & 0.80\%\\ 
M 7.8 & April 26, 2001 & 9.20 & 1.21 & 1.04 & 1.09 & 1.10 & 0.40 & 0.05 & 0.07 & 0.06 & 68.72\% & 77.54\% & 94.92\% & 98.93\%\\ 
M 8.0 & July 25, 2000 & 2.80 & 0.76 & 0.81 & 0.80 & 0.74 & 0.26 & 0.04 & 0.03 & 0.03 & 16.04\% & 0.00\% & 0.00\% & 0.00\%\\ 
M 8.2 & April 10, 2002 & 4.90 & 0.88 & 0.93 & 0.95 & 0.96 & 0.30 & 0.04 & 0.07 & 0.08 & 27.81\% & 5.08\% & 21.12\% & 24.06\%\\ 
M 9.1 & Nov. 8, 2001 & 1.70 & 1.22 & 1.10 & 1.11 & 1.11 & 0.47 & 0.10 & 0.12 & 0.13 & 62.83\% & 82.62\% & 82.89\% & 79.68\%\\ 
M 9.5 & Sep. 9, 2001 & 2.20 & 1.06 & 0.97 & 0.99 & 1.00 & 0.37 & 0.07 & 0.10 & 0.11 & 49.73\% & 28.88\% & 39.84\% & 43.58\%\\ 
M 9.6 & Nov. 20, 2003 & 6.00 & 1.06 & 1.00 & 1.00 & 1.00 & 0.37 & 0.06 & 0.08 & 0.10 & 50.80\% & 54.01\% & 49.73\% & 48.40\%\\
X 1.2 & June 7, 2000 & 14.00 & 1.15 & 1.06 & 1.12 & 1.18 & 0.38 & 0.05 & 0.07 & 0.07 & 63.37\% & 91.71\% & 98.40\% & 100.00\%\\
X 1.2 & June 23, 2001 & 2.60 & 0.89 & 1.06 & 1.00 & 0.95 & 0.45 & 0.09 & 0.10 & 0.09 & 32.62\% & 79.41\% & 54.55\% & 32.89\%\\
X 1.3 & May 27, 2003 & 7.10 & 0.86 & 0.92 & 0.94 & 0.97 & 0.31 & 0.03 & 0.07 & 0.09 & 23.80\% & 0.80\% & 17.65\% & 30.21\%\\ 
X 1.7 & March 29, 2001 & 22.00 & 1.02 & 1.06 & 1.05 & 1.04 & 0.32 & 0.10 & 0.09 & 0.08 & 47.59\% & 75.13\% & 72.73\% & 68.45\%\\
X 1.8 & July 18, 2002 & 5.60 & 1.08 & 1.04 & 1.10 & 1.13 & 0.40 & 0.11 & 0.17 & 0.17 & 48.66\% & 57.49\% & 74.87\% & 79.95\%\\
X 1.9 & July 12, 2000 & 14.00 & 0.84 & 0.81 & 0.77 & 0.71 & 0.29 & 0.05 & 0.05 & 0.05 & 22.99\% & 0.00\% & 0.00\%  & 0.00\%\\
X 2.3 & Nov. 24, 2000 & 16.00 & 1.24 & 0.98 & 0.98 & 0.99 & 0.42 & 0.05 & 0.05 & 0.04 & 68.98\% & 37.43\% & 32.09\% & 39.57\%\\
X 2.3 & April 10, 2001 & 30.00 & 0.71 & 0.67 & 0.64 & 0.63 & 0.29 & 0.04 & 0.05 & 0.04 & 14.71\% & 0.00\% & 0.00\% & 0.00\%\\
X 2.6 & Sep. 24, 2001 & 63.00 & 1.31 & 1.10 & 1.12 & 1.12 & 0.64 & 0.13 & 0.16 & 0.16 & 64.97\% & 75.40\% & 75.94\% & 76.47\%\\
X 3.0 & July 15, 2002 & 14.00 & 1.37 & 1.08 & 1.06 & 1.05 & 0.61 & 0.06 & 0.08 & 0.09 & 70.32\% & 96.52\% & 82.62\% & 73.26\%\\
X 4.0 & Nov. 26, 2000 & 28.00 & 0.91 & 1.00 & 1.00 & 0.98 & 0.27 & 0.05 & 0.05 & 0.04 & 32.09\% & 42.78\% & 47.06\% & 23.53\%\\
X 5.3 & August 25, 2001 & 82.00 & 1.09 & 0.99 & 0.98 & 1.02 & 0.66 & 0.08 & 0.12 & 0.13 & 44.92\% & 37.97\% & 36.90\% & 49.73\%\\
X 5.6 & April 6, 2001 & 41.00 & 0.98 & 0.81 & 0.82 & 0.78 & 0.46 & 0.06 & 0.05 & 0.04 & 37.43\% & 0.27\% & 0.27\% & 0.00\%\\
X 5.7 & July 14, 2000 & 75.00 & 1.30 & 1.14 & 1.23 & 1.41 & 0.50 & 0.04 & 0.07 & 0.11 & 68.72\% & 100.00\% & 100.00\% & 100.00\%\\
X 6.2 & Dec. 13, 2001 & 25.00 & 0.72 & 0.65 & 0.65 & 0.66 & 0.26 & 0.05 & 0.06 & 0.08 & 13.37\% & 0.00\% & 0.00\% & 0.27\%\\
X 17.2 & Oct. 28, 2003 & 180.00 & 0.66 & 0.72 & 0.74 & 0.74 & 0.23 & 0.06 & 0.07 & 0.09 & 7.49\% & 0.27\% & 0.53\% &  1.34\%\\
\bottomrule
\end{tabular}
\begin{tablenotes}
\item[a] Average value of power ratio above the acoustic cutoff frequency at 5.3 mHz.
\item[b] Standard deviation of power ratio above 5.3 mHz.
\item[c] Percentage of values greater than unity above 5.3 mHz. Percentages of zero indicate that there was no increase in power in the high-frequency regime of the power spectrum after the occurrence of the flare.
\item[d] $\ell_1$, $\ell_2$, $\ell_3$, and $\ell_4$ are representative of degree ranges $\ell$ = 0 to $\ell$ = 50, $\ell$ = 51 to $\ell$ = 100, $\ell$ = 101 to $\ell$ = 150, and $\ell$ = 151 to $\ell$ = 200, respectively.
\end{tablenotes}
\end{threeparttable}
\end{table}
\end{landscape} 

\subsection{Power as a function of frequency and time}
\begin{figure}[h!]
\centerline{
\includegraphics[width=5.0 in]{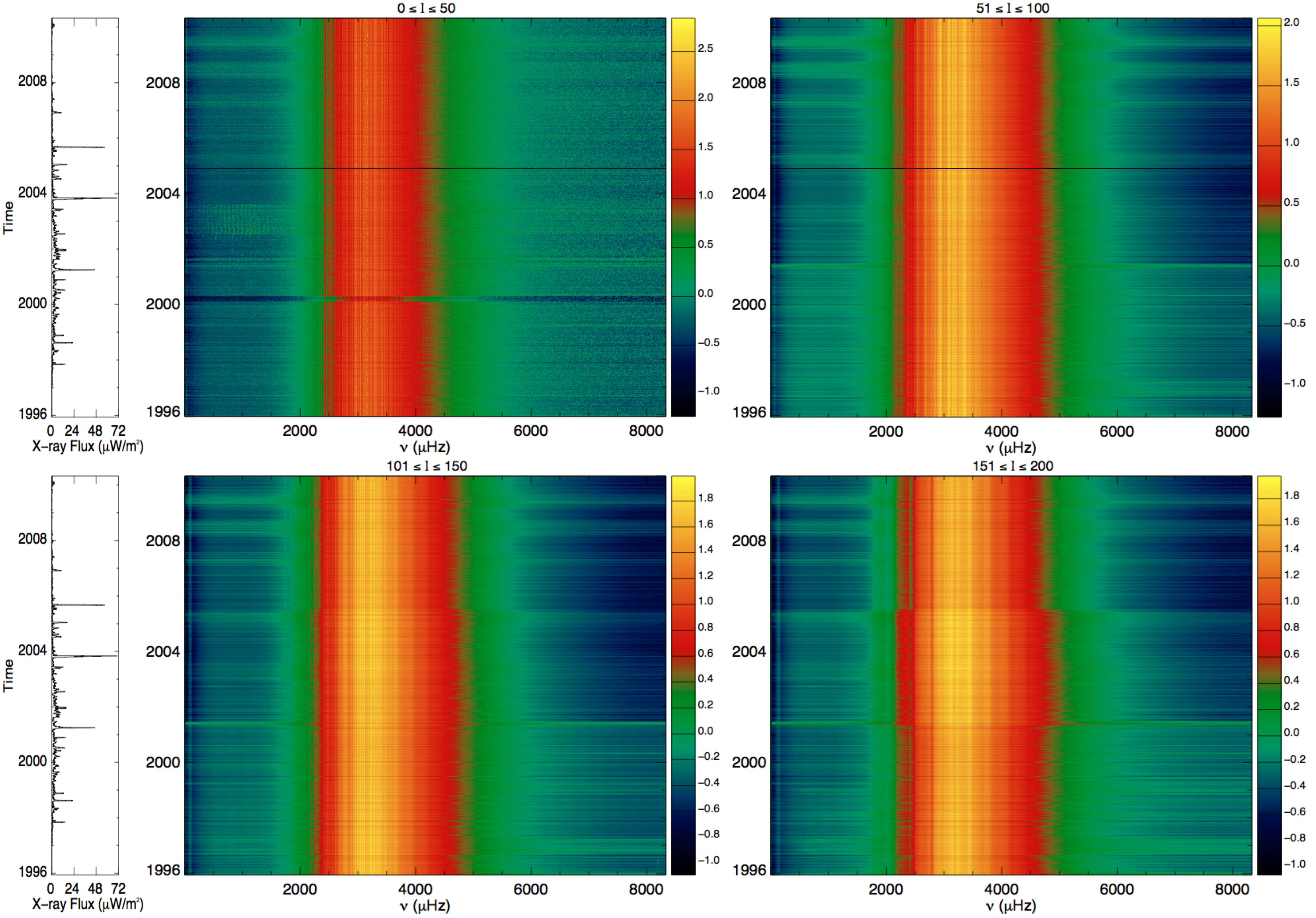}
}
\caption{The color images show the averaged oscillation power spectrum as a function of year and frequency $\nu$ for the four ranges of $\ell$.
The color scale shows the magnitude of the power and is logarithmically scaled. Note that the color scales are different for each of the images. 
The red/yellow vertical bars starting around 2 mHz and ending around 4.5 mHz in the color images are the $p$-mode oscillations. The sharp decrease in power seen in the top left panel around early 2000 and late 2004 as well as in the top right panel around late 2004 is due to outlier removal. 
The two narrow panels on the left of the figure show the average soft X-ray flux as a function of time.  The large horizontal spikes correspond to major solar flare occurrences. 
}
\label{Fig: 7}
\end{figure}

Figure~\ref{Fig: 7} shows the average $p$-mode power as function of time and frequency in our four ranges of spherical harmonic degree $\ell$. The prominent red/yellow bands are indicative of the usual $p$ modes, which have typical frequencies between 2.2 and 5 mHz. The left side of Figure~\ref{Fig: 7} shows the GOES soft X-ray flux as a function of time, averaged over the same 7.5 day periods offset by 18 hours that were used for computing the power spectra. Note that the four color plots have different scales.

\begin{figure}[h!]
\centerline{
\includegraphics[width=5.0 in]{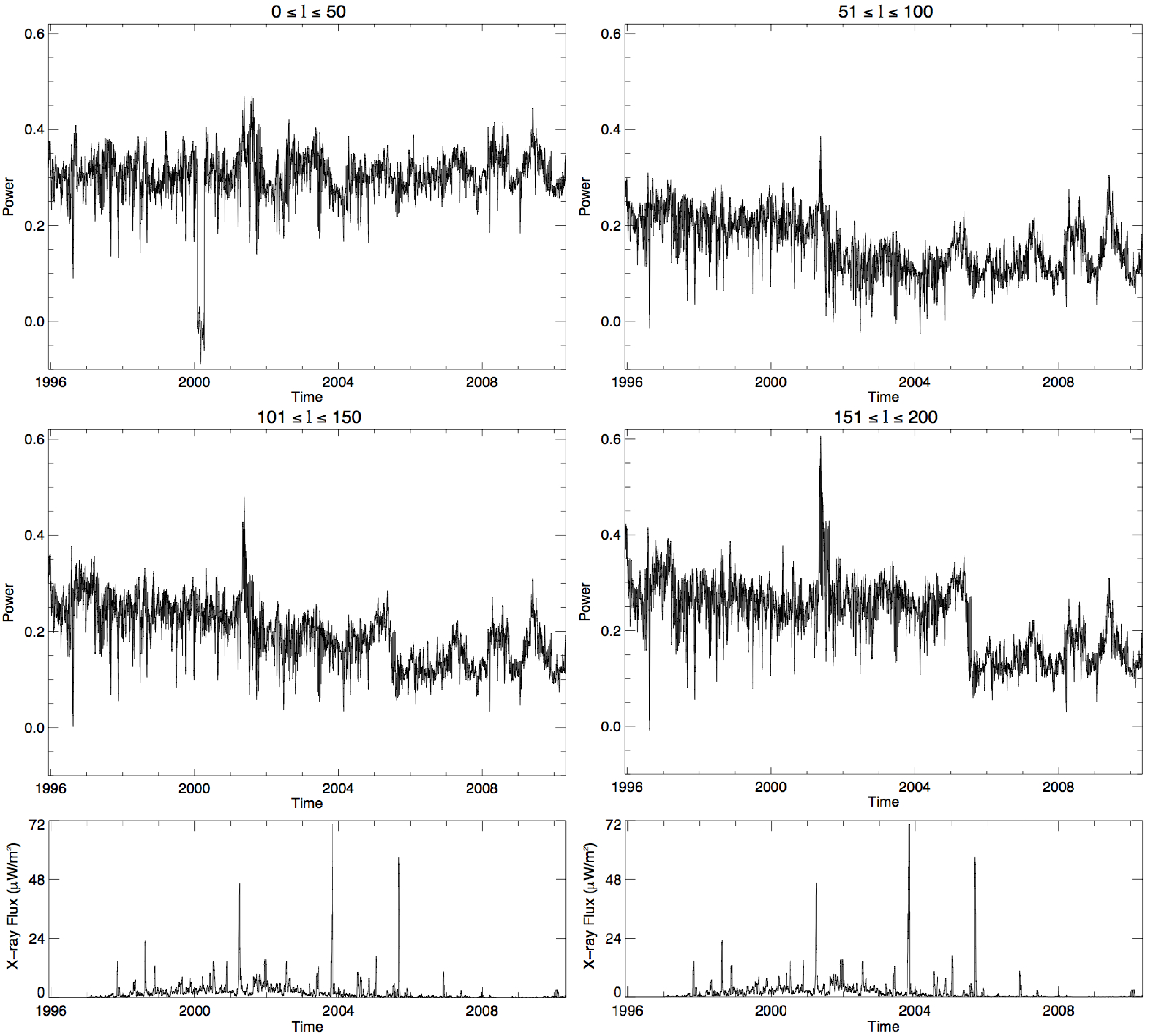}
}
\caption{The temporal variation of $p$-mode power averaged over all frequencies. The X-ray flux is shown the in bottom left and right panels. There does not appear to be any consistent relationship between X-ray flux and acoustic power. }
\label{Fig: 8}
\end{figure}

A comparison of the location of the peaks in the soft X-ray flux with the localized variations in the oscillation power spectra does not show any obvious correlation between the X-ray flux and enhancements of oscillation power. The lack of correlation can also be seen in Figure~\ref{Fig: 8}, where the data in Figure~\ref{Fig: 7} have been averaged over all frequencies, producing the temporal evolution of the oscillation power in our four ranges of $\ell$. 

Both Figures~\ref{Fig: 7} and~\ref{Fig: 8} show a number of interesting features. The large drop in power in the $\ell_1$ range in early 2000 is very likely a data processing artifact, as is the similar drop evident in Figure~\ref{Fig: 7} in the $\ell_1$ and $\ell_2$ ranges in late 2004. A significant number of outliers were discarded from the power spectra at these times.

There is a sharp local increase in power in all ranges of $\ell$ in April 2001, which follows a series of three major flares within one month. These flares were an X1.7 on March 29, 2001; a X2.3 on April 10, 2001, and an M7.8 on April 26, 2001. It is possible that this sequence may have indeed amplified the strength of the oscillations. Finally, Figure~\ref{Fig: 8} shows that there is an abrupt and then persistent change in the overall power level of the oscillations in the $\ell_3$ and $\ell_4$ ranges in mid 2005. The change is also apparent in Figure~\ref{Fig: 7} as a change in the width in frequency of the $p$-mode band.

The origin of this change is most likely due to a revision of the method used to reconcile the angular orientation of the solar images around the GONG network. Since it is impossible to perfectly align the instruments north-south on the earth, the apparent position of the solar rotation axis varies by a few tenths of a degree between the six GONG sites, and this discrepancy changes over the course of the year. An angular change in the rotation axis orientation redistributes the power attributed to a given oscillation mode with specific values of $\ell$ and $m$ into other values of $\ell$ and $m$, and this introduces errors into measurements of the frequency and amplitude of the modes (\citeauthor{Kennedy97} \citeyear{Kennedy97}; \citeauthor{Kennedy98} \citeyear{Kennedy98}). These errors are larger at higher values of $\ell$, and also increase with an increase in the rotation axis discrepancy, so considerable effort has been made to reduce the site-to-site angular discrepancies in the GONG data and to determine the true orientation of the SunÕs rotation axis. It is estimated that these efforts have decreased the systematic error in the axis orientation by an order of magnitude, down to 0.02$^{\circ}$. In 2005, the GONG Project introduced a scheduled drift scan observation at local noon around the network that substantially improved the accuracy of the orientation measurement. This is the source of the amplitude drop in 2005 seen in Figure~\ref{Fig: 8}, which shows no drop for the lowest range of $\ell$, and an increase in the drop as $\ell$ increases. This is consistent with the known properties of the effect.

\begin{figure}[h!]
\centerline{\includegraphics[width=5.0 in]{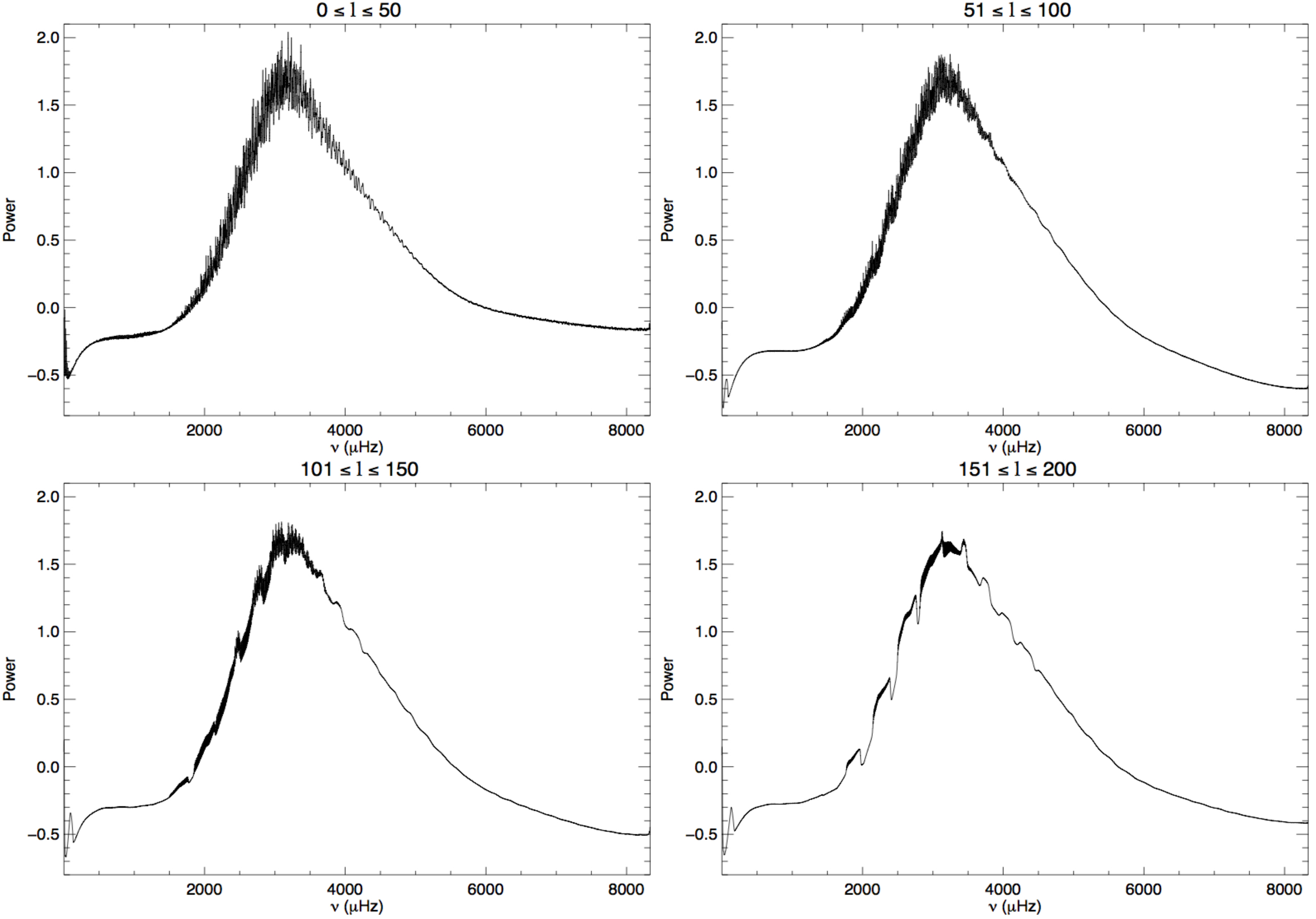}
}
\caption{The frequency variation of $p$-mode power averaged over all time.}
\label{Fig: 9}
\end{figure}

Figure~\ref{Fig: 9} shows the corresponding time averaged power spectrum as a function of frequency for the four ranges in $\ell$. We see that the acoustic power at frequencies higher than the acoustic cut-off frequency around 5.3 mHz are greatest in the $\ell_1$ range ($\ell$ = 0  to $\ell$ = 50) (top left panel). This suggests that the solar processes that excite the high-frequency oscillations do so mainly at low and intermediate degrees.
The time-averaged spectra for the $\ell_3$ and $\ell_4$ ranges in Figure~\ref{Fig: 9} display some notch-like jumps, which are also the result of the change in the determination of the solar axis orientation as discussed above. As before, Figure~\ref{Fig: 9} shows that the magnitude of the change increases with the range of $\ell$ and, in addition,  that the power redistribution occurs at frequencies that also depend on the range of $\ell$.

\section{Discussion}
\noindent 
In this study, we have searched for evidence that large and energetic solar flares can change the amplitude of the global oscillations of the Sun, as postulated by \citeauthor{Wolff72} (\citeyear{Wolff72}), and apparently observed in SOHO data by \citeauthor{Karoff08} (\citeyear{Karoff08}). We find no evidence for this process in the GONG data. Ratios of $m$- and $\ell$-averaged spectra obtained before and after 31 flares of class M5 or higher essentially show that a decrease in the mode amplitude is just as likely as an increase, and images of the mode power as a function of time and frequency do not show any significant variations in time that can be associated with flares. Our results do show variations that are most likely the result of the stochastic excitation of the solar $p$ modes. There are some puzzling facets of this variation, such as the relatively large amplitude variations at low spherical harmonic degrees below $\ell = 50$, compared to the variations at higher degrees. The variance of the mode amplitude is also observed to be higher at all $\ell$ above the acoustic cutoff frequency of 5.3 mHz. This may be due to the lack of trapping of waves with frequencies higher than the acoustic cutoff, which could allow the rapidly evolving solar atmosphere to have a larger influence on the coherence of the waves compared to the relatively quiescent interior.

With the exception of \citeauthor{Kumar11} (\citeyear{Kumar11}), other studies have not detected flare-related changes in the global mode amplitudes as observed by GONG (\citeauthor{Chakraborty11} \citeyear{Chakraborty11}; \citeauthor{Ambastha06} \citeyear{Ambastha06}), but studies using data from instruments on SOHO generally have detected such changes (\citeauthor{Karoff08}, \citeyear{Karoff08}; \citeauthor{Kumar09}, \citeyear{Kumar09}; \citeauthor{Kumar10}, \citeyear{Kumar10};  \citeauthor{Chakraborty11}, \citeyear{Chakraborty11}). There are at least five possible explanations for this discrepancy. One is that the noise level of the GONG observations is substantially higher than the noise in the SOHO instruments, but there have been many comparisons between the data sets since 1995 (see, e.g., \citeauthor{Komm98}, \citeyear{Komm98}), and there is no indication that this is the case. A second possibility is that all of the instruments used by various studies of this question have different distributions of the spatial leaks discussed in Sec. 2. These leaks may somehow mask the amplitude variations associated with flares in the GONG data. The third possibility is that the flare effects are restricted to only the very low-degree modes, and that averaging over a range in $\ell$ of 0 to 50 has masked the effects. On the other hand, it would be surprising if a very localized impulse such as flare would only affect very low spatial frequencies, since the spectrum of an impulse in the Fourier domain is essentially white, with all spatial frequencies present.
The fourth possibility is that the SOHO instruments are directly affected by the flares since they are located in space and exposed to the excess particle emission generated by these events. Ground-based instruments are shielded from this by the Earth's atmosphere. It is possible that the apparent changes in the oscillation amplitudes are due to increased noise when charged particles hit the space-based detectors. A test of this would be to search for flare effects using BiSON data, which is another ground-based network.
The fifth possibility may be related to the way in which the data have been reduced and analyzed. An analysis of MDI, GOLF, and VIRGO data using the methodology of this analysis would be an ideal way to test for discrepancies due to data processing. Although the aforementioned analyses are outside the scope of this effort, it is possible that they will be done for a future paper.

The results also raise some additional questions about the interaction of the flares and the global oscillations. The flare associated with the largest increase in amplitude in Table 1 is the M5.9 flare on October 5, 2002. The power of the oscillations basically doubled for this event at all degrees, even though it was a relatively weak event. Why? Was this simply a coincidence between the flare and a strong random fluctuation at all values of $\ell$, or was there some special aspect of the event, such as its location on the solar surface? Does the location or time of an event play a role in how effective a flare may or may not be in modifying the oscillations? Finally, does a sequence of flares, such as occurred in March and April of 2001, have a cumulative effect on the modes? If so, how often and how strong do the flares have to be in order to cause such an effect?
With the advent of asteroseismology, we may be able to address these questions using statistical studies on data sets from Kepler and the Stellar Observations Network Group (SONG).

%Figures 
\newpage

\newpage

\newpage

\newpage

\newpage

\newpage

\newpage

\newpage

\newpage

%% Table
%
% \begin{table}
% \caption{}%\label{tbl:?}
% \begin{tabular}{}     3
% \hline
% \multicolumn{2}{c}{<>}
% <data>
% \hline
% \end{tabular}
% \end{table}

%%%%%%%%%%%%%%%%%%%%%%%%%%%%%%%%%%%%%%%%%%%%%%%%%%%%%%%%%%%%%%%%%%%%%%%%%%%
%% Appendix
%
% \appendix   

%Acknowledgements
\begin{acks}
This work utilizes data obtained by the GONG program, managed by the National Solar Observatory (NSO), which is operated by AURA, Inc. under a cooperative agreement with the National Science Foundation (NSF). The data was acquired by instruments operated by the Big Bear Solar Observatory, High Altitude Observatory, Learmonth Solar Observatory, Udaipur Solar Observatory, Instituto de Astrof«õsica de Canarias, and Cerro Tololo Interamerican Observatory. This work was made possible by NSF PAARE grant AST-0849736, which supports the Fisk-Vanderbilt Masters-to-PhD Bridge Program and the partnership with NSO.  K.G.S. acknowledges support from a Martin Luther King Visiting Professorship at the Massachusetts Institute of Technology.
\end{acks}

%Bibliography
%
% Using BibTeX
%
\bibliographystyle{spr-mp-sola}
% %\bibliographystyle{spr-mp-sola-cnd} %% Alternative style: no title, no concluding page
\bibliography{ReferenceList}
% Checking: look if the file containing the ``\bibitem'' exits
 %           so check if the .bbl file exist (bibTeX compilation)
\IfFileExists{\jobname.bbl}{} {\typeout{}
\typeout{****************************************************}
\typeout{****************************************************}
\typeout{** Please run "bibtex \jobname" to obtain} \typeout{**
the bibliography and then re-run LaTeX} \typeout{** twice to fix
the references !}
\typeout{****************************************************}
\typeout{****************************************************}
\typeout{}}
% Without BibTeX 
% \begin{thebibliography}{}
% \bibitem[\protect\citeauthoryear{Author}{Year}]{key}
%   <bibliographical entry>
%
% \bibitem[\protect\citeauthoryear{}{}]{}
%   
%  
% \end{thebibliography}

\end{article} 
\end{document}